\documentclass[
    ,final            % use final for the camera ready runs
    ,numberedheadings % uncomment this option for numbered sections
  ]
  {aipproc}

\layoutstyle{6x9}

\begin{document}

\title{The Second Stellar Spectrum and the non-LTE Problem of the 2nd Kind\footnote{Invited review paper published in {\em Recent Directions in Astrophysical Quantitative Spectroscopy and Radiation Hydrodynamics}. Proceedings of the International Conference in Honor of Dimitri Mihalas for His Lifetime Scientific Contributions on the Occasion of His 70th Birthday. Editors: I. Hubeny, J. M. Stone, K. McGregor and K. Werner. {\em American Institute of Physics}. AIP Conference Proceedings 1171, pp. 27 -- 42, (2009).}}


\author{Javier Trujillo Bueno}{
  address={Instituto de Astrof\'\i sica de Canarias; 38205 La Laguna; Tenerife; Spain},altaddress={Consejo Superior de Investigaciones Cient\'\i ficas (Spain)} 
}

\begin{abstract}
This paper presents an overview of the radiative transfer problem of calculating the
spectral line intensity and polarization that emerges from a (generally magnetized) 
astrophysical plasma composed of atoms and molecules 
whose excitation state is significantly influenced by radiative transitions 
produced by an anisotropic radiation field. The numerical solution of this non-LTE problem 
of the 2nd kind is facilitating the physical understanding of the second solar spectrum and the 
exploration of the complex magnetism of the extended solar atmosphere, but much more could be learned if high-sensitivity polarimeters were developed also for the present generation of night-time telescopes. Interestingly, I find that the population ratio between the levels of some resonance line transitions can be efficiently modulated by the inclination of a weak magnetic field when the anisotropy of the incident radiation is significant, something that could provide a new diagnostic tool in astrophysics. 
\end{abstract}

\maketitle

%%%%%%%%%%%%%%%%%%%%%%%%%%%%%%%%%%%%%%%%%%%%
%% MAINMATTER
%%%%%%%%%%%%%%%%%%%%%%%%%%%%%%%%%%%%%%%%%%%%

\section{Introduction}

The non-LTE problem of the 1st kind is that 
described in Mihalas's (1978) book \citep{Mihalas-1978}: it consists in calculating, at 
each spatial point of the astrophysical plasma model under consideration, the values of the atomic level 
populations that are consistent with the intensity of the radiation field generated within the medium. This 
requires solving jointly the radiative transfer (RT) equations for the specific intensities and the statistical 
equilibrium equations for the level populations. Nowadays, this non-local and non-linear problem is routinely solved via the application of ``clever'' iterative schemes, where everything goes as simply as in the $\Lambda$-iteration method, but for which the convergence rate is extremely high 
\citep[e.g., the reviews by][]{Hubeny-2003,Trujillo-Bueno-2003}. 
The success of this international meeting to honor Prof. Dimitri Mihalas on the occasion of his 70th birthday is a clear demonstration that the numerical solution of this multilevel radiative transfer problem has enabled many important advances to be made in 20th century astrophysics, including the modeling of the Fraunhofer spectrum and the development of the field of astrophysical spectroscopy. 

In contrast, the non-LTE problem of the 2nd kind 
is still rather unfamiliar to astrophysicists. This term was introduced by Landi Degl'Innocenti \citep{Landi-1987} to refer to the formidable numerical problem that implies calculating, at each spatial 
point of the (generally magnetized) astrophysical plasma model under consideration, the values of the $(2J+1)^2$ elements of the atomic density matrix corresponding to each atomic level of total angular momentum $J$, which quantify its overall population, the population imbalances between its magnetic sublevels, and the quantum coherences between each pair of them. The values of such density-matrix elements have to be consistent with the intensity, polarization and symmetry properties of the 
radiation field generated within the medium. This requires solving jointly the RT 
equations for the Stokes parameters and the statistical equilibrium equations for the elements of the atomic density matrix\footnote{Note that the first Stokes parameter, $I$, is the specific intensity at a given wavelength, while Stokes $Q$ represents  the intensity difference between linear polarization parallel and perpendicular to a given reference direction (e.g., the direction perpendicular to the straight line joining the solar disk center with the observed point). Stokes $U$ would be then the intensity difference between linear polarization at $+45^{\circ}$ and $-45^{\circ}$ with respect to the chosen reference direction, where positive angles are measured counterclockwise by an observer facing the radiation. Stokes $V$ is the intensity difference between right-handed and left-handed circular polarization.}.  
Although much remains to be done, the numerical solution of the non-LTE problem of the 2nd kind is facilitating the physical understanding and modeling of the so-called {\em second solar spectrum}, as well as the development of the field of astrophysical spectropolarimetry.

As clarified below, the second solar spectrum is the observational signature of radiatively induced population imbalances and quantum coherences in the atoms and molecules of the solar atmosphere. Magnetic fields produce partial decoherence via the Hanle effect, giving rise to fascinating observable effects in the emergent spectral line polarization. Interestingly, these effects allow us to ``see'' magnetic fields to which the Zeeman effect is blind within the limitations of the available instrumentation \citep[e.g.,][]{Trujillo-Bueno-2004}. In fact, the Zeeman effect is of limited practical interest for the exploration of magnetic fields in hot (e.g., chromospheric and coronal) plasmas because the polarization that it induces in a spectral line scales with the ratio between the Zeeman splitting and the Doppler width, and this ratio turns out to be particularly small in such plasmas. Moreover, the polarization of the Zeeman effect as a diagnostic tool is blind to magnetic fields that are randomly oriented on scales too small to be resolved. 
We may thus define ``the Sun's hidden magnetism'' as all the magnetic fields of the extended solar atmosphere that are impossible to diagnose via the consideration of the Zeeman effect alone.

\begin{figure}
   \includegraphics[height=.40\textheight]{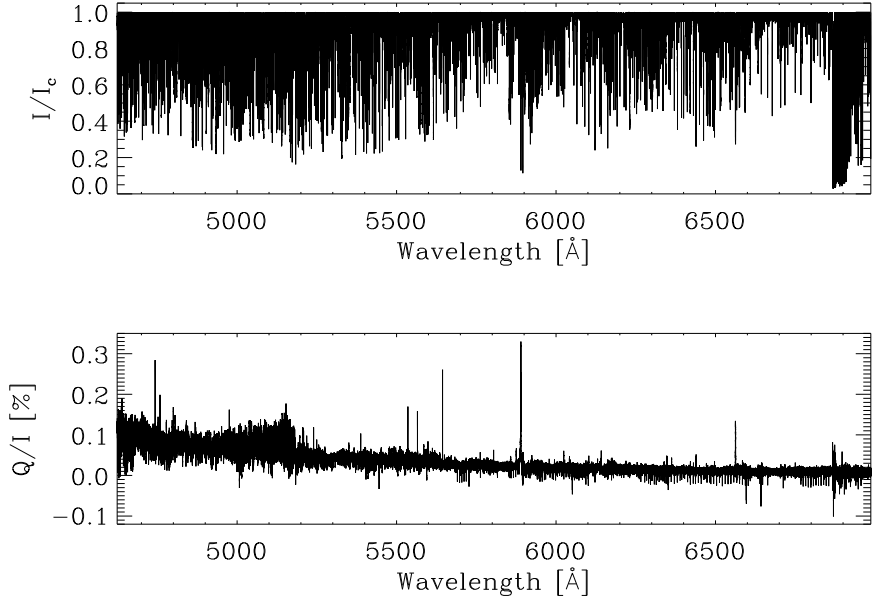}
  \caption{The visible Fraunhofer spectrum (displayed as the normalized intensity of the continuum) versus the second solar spectrum (displayed as the fractional linear polarization, $Q/I$). The positive direction for the Stokes $Q$ parameter is the parallel to the observed solar limb. Extracted from \citep{Gandorfer-2000}.}
\end{figure} 

\section{The Second Stellar Spectrum} 
 
Figure 1 contrasts the Fraunhofer spectrum (i.e., the visible intensity spectrum of the Sun with its multitude of absorption lines) with the so-called second solar spectrum, which is the linearly polarized spectrum observed  by Stenflo and collaborators in ``quiet'' regions of the solar atmosphere 
\citep[see][]{Stenflo-Keller-1997,Stenflo-2000,Gandorfer-2000,Gandorfer-2002}. The fractional linear polarization ($Q/I$) of the bottom panel shows a number of conspicuous peaks on top of an {\em aparently noisy} $Q/I$ background. Such $Q/I$ peaks  correspond to $\lambda$6562.8 (H$_{\alpha}$), $\lambda$5890 (Na {\sc i} D$_2$), $\lambda$5644.14 (Ti {\sc i}), $\lambda$5565.48 (Ti {\sc i}), $\lambda$5535.51 (Ba {\sc i}), $\lambda$5167.32 (Mg {\sc i}), $\lambda$4758.12 (Ti {\sc i}) and $\lambda$4742.79 (Ti {\sc i}). For example, the left panels of Fig. 2 show an enlarged portion around the wavelengths of the 
sodium doublet. Similarly rich $Q/I$ spectral structure is found in many other spectral regions, especially at wavelengths smaller than 5300 \AA\ (e.g., see the right panels of Fig. 2, which show the $Q/I$ profiles produced by Y {\sc ii}, Cr {\sc i} and Fe {\sc i}). Clearly, the second solar spectrum has a structural richness that often exceeds that of the ordinary intensity spectrum on which most of our astrophysics is based. It is also of interest to note that the linear polarization amplitudes tend to increase towards the blue and near-UV spectral regions. For example, between 3910 \AA\ and 4630 \AA\ we find spectral lines with $Q/I$ amplitudes of the order of $1\%$, such as Ca {\sc i} $\lambda$4227, Sr {\sc ii} $\lambda$4078, Ba {\sc ii} $\lambda$4554 and Sr {\sc i} $\lambda$4607 \citep[see][]{Gandorfer-2002}.

\begin{figure}
   \includegraphics[height=.242\textheight]{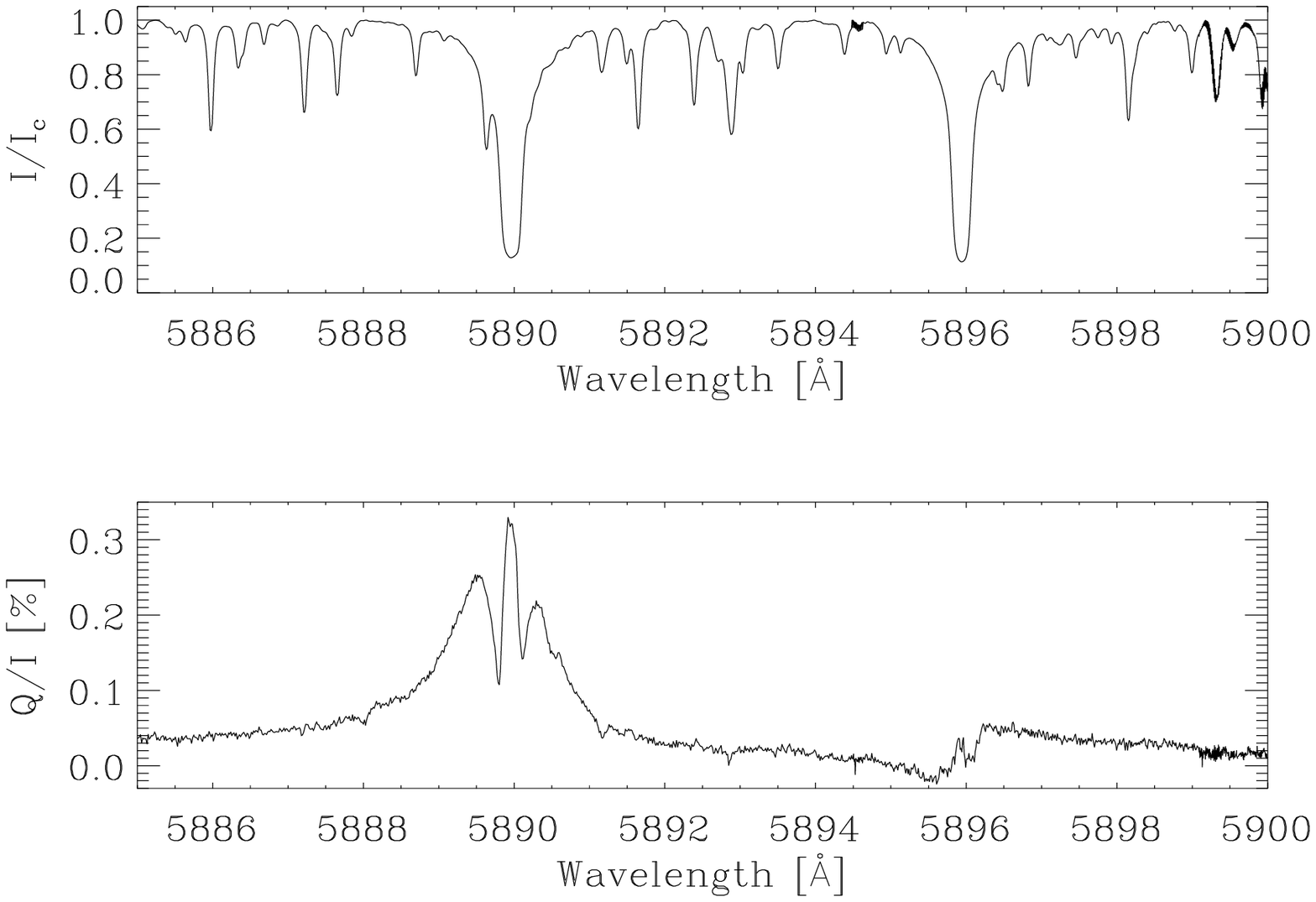}
   \includegraphics[height=.242\textheight]{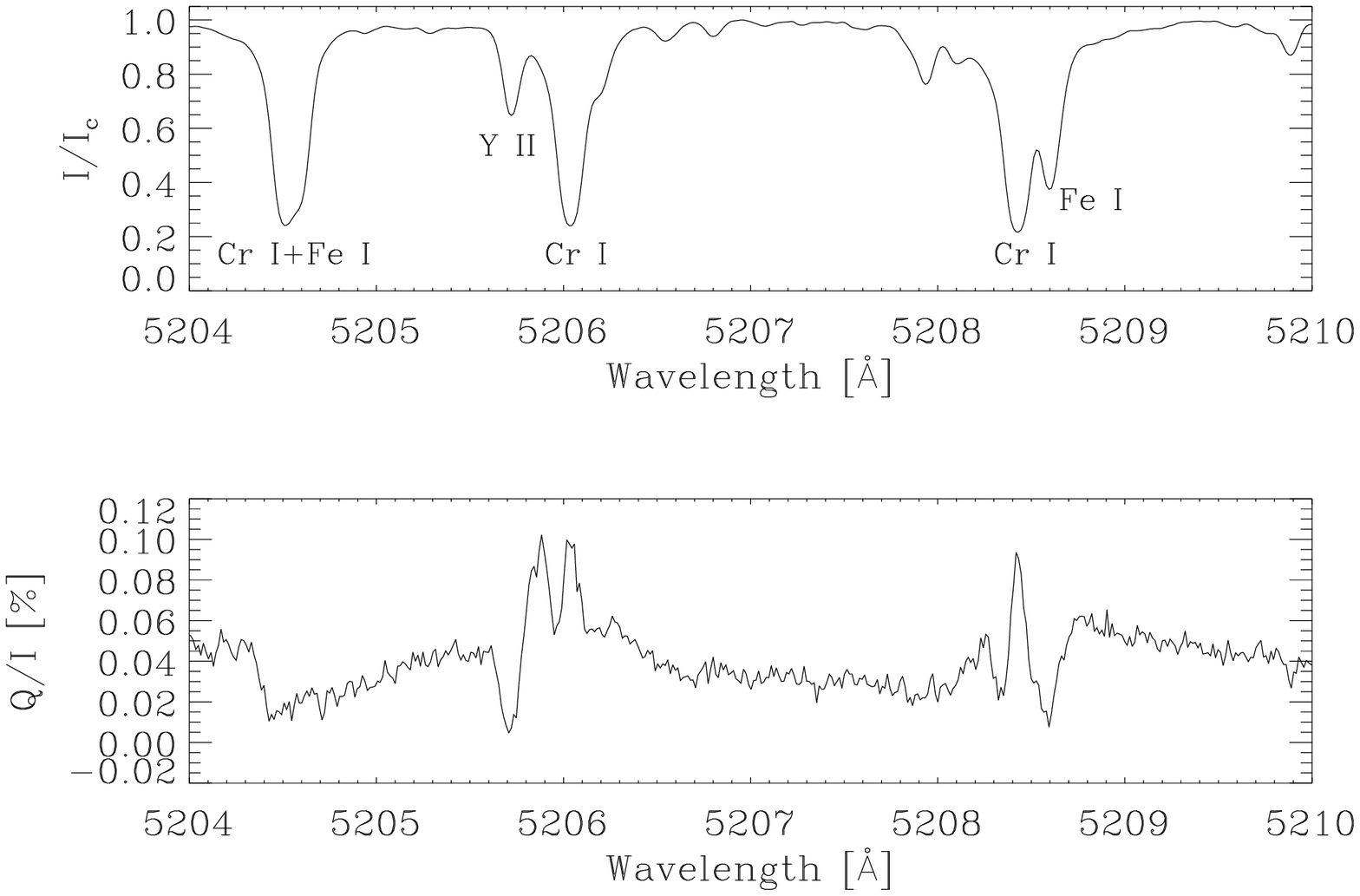}
   \caption{The two left panels are enlarged portions of Fig. 1 showing the complex $Q/I$ structure across the sodium doublet, while the two right panels focus on a very interesting spectral region showing lines of Y {\sc ii}, Cr {\sc i} and Fe {\sc i}.}
\end{figure}  

As is well known, the polarization of the Zeeman effect is due to the wavelength shifts between the $\pi$ (${\Delta}M=M_u-M_l=0$) and $\sigma_{b,r}$ (${\Delta}M={\pm}1$) transitions\footnote{Such wavelength shifts are due to the presence of a magnetic field, which causes the atomic energy levels to split into different magnetic
sublevels characterized by their magnetic quantum number $M$.}. The typical observational signature of the circular polarization produced by the longitudinal Zeeman effect is an antisymmetric Stokes $V(\lambda)$ profile whose amplitude scales with the ratio, ${\cal R}$, between the Zeeman splitting and the Doppler broadened line width. The linear polarization amplitudes of the transverse Zeeman effect scale instead as ${\cal R}^2$ and its typical observational signatures are symmetric Stokes $Q(\lambda)$ and $U(\lambda)$ profiles with their wing lobes of opposite sign to the line center one. 

As seen in the bottom panels of Fig. 2, the linear polarization signals of the second solar spectrum 
have nothing to do with the transverse Zeeman effect. This can also be seen 
in Fig. 3, which shows a high-sensitivity spectropolarimetric observation of the solar limb atmosphere in
the strongest (8542~\AA) chromospheric line of the Ca~{\sc ii} triplet. The
observed Stokes $V/I$ profiles are clearly caused by the longitudinal
Zeeman effect, but the Stokes $Q/I$ and $U/I$ signals have  
Gaussian-like shapes at almost all spatial positions along the direction 
of the spectrograph's slit. Interestingly, while the Stokes
$Q/I$ signal changes its amplitude but remains always positive along
that spatial direction, the sign of the Stokes $U/I$ signal
fluctuates.

The shapes of the $Q/I$ profiles of the second solar spectrum are not always Gaussian-like, however. A classification of the various $Q/I$ shapes of atomic lines has been recently proposed by \citep{Belluzzi-2009}. Strong resonance lines tend to show a triple peak structure, as is the case with the Na {\sc i} D$_2$ line shown in the bottom left panel of Figure 2. However, the shape of the observed $Q/I$ and $U/I$ profiles may also show spatial variability, as illustrated in Fig. 4 for the chromospheric K-line of Ca {\sc ii}. Note that in this near-UV line there is no measurable hint of the longitudinal Zeeman effect, in contrast with the case of the near-IR line of Fig. 3 whose line center Stokes $Q/I$ and $U/I$ signals also originate in the bulk of the solar chromosphere.

\begin{figure}
  \includegraphics[height=.12\textheight]{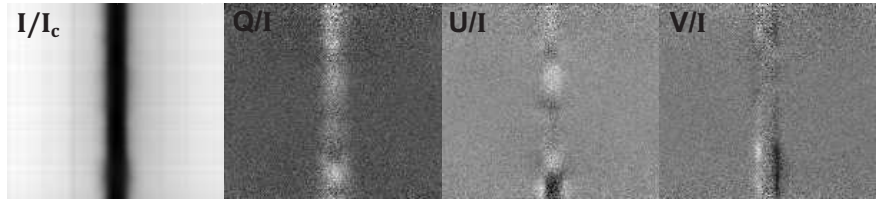}
  \caption{Spectropolarimetric observation of the
  Ca~{\sc ii} 8542~\AA\ line in a very quiet region close to the solar
  limb, obtained with the Z\"urich Imaging Polarimeter (ZIMPOL) 
  at the Franco-Italian telescope THEMIS of the Observatorio del Teide (Tenerife; Spain). 
  The reference direction for Stokes
  $Q$ is the tangent to the closest limb. From \citep{Trujillo-etal-2009}.}
\end{figure}

\begin{figure}
  \includegraphics[height=.35\textheight]{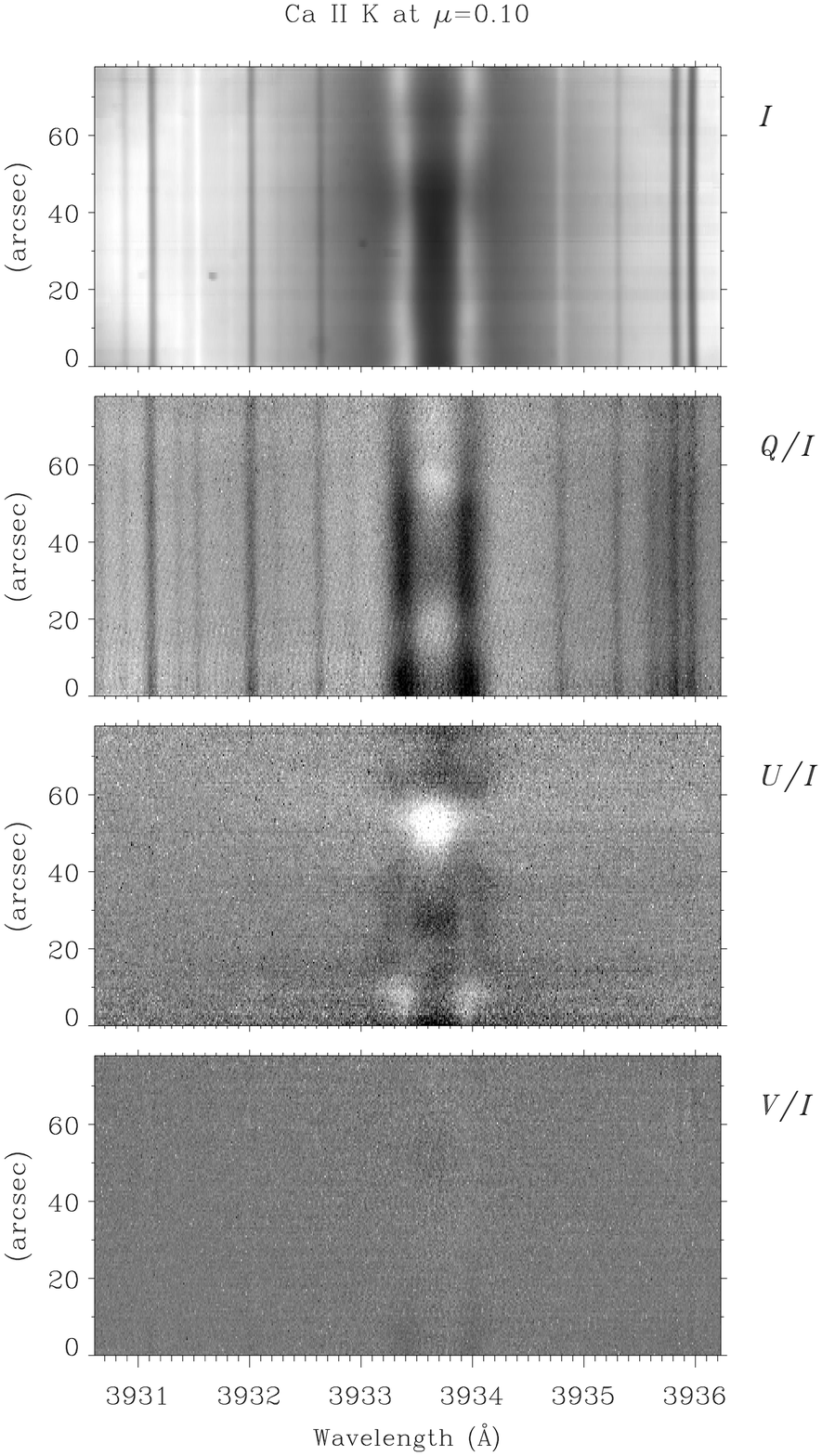}
  \includegraphics[height=.35\textheight]{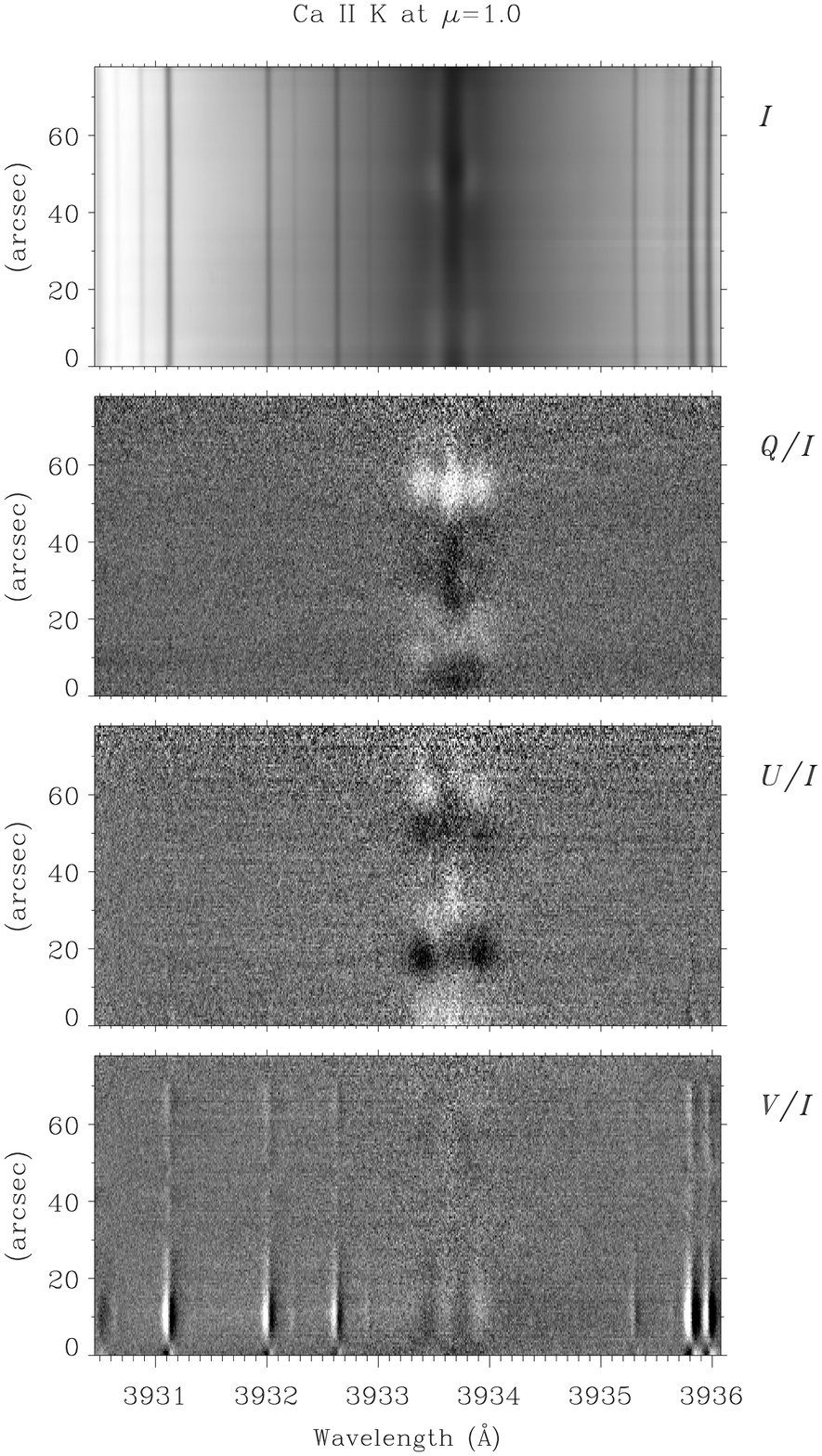}
  \caption{Spectropolarimetric observation of the K-line of Ca~{\sc ii} in quiet regions of the solar 
  atmosphere, obtained with ZIMPOL at the McMath Telescope in Kitt Peak Observatory (USA). 
  Left panel: observation close to the solar limb. Right panel: disk center observation. 
  From \citep{Stenflo-2006}.}
\end{figure} 
 
\section{The physical origin of the 2nd stellar spectrum}

If the second solar spectrum is not caused by the Zeeman effect, then what could its physical origin be?

\subsection{Atomic level polarization and anisotropic radiation pumping}

In stellar atmospheres there is
another fundamental mechanism, besides the Zeeman effect, producing polarization in spectral
lines. There, where the emitted radiation can escape through the
stellar surface, the atoms are illuminated by an anisotropic
radiation field. The ensuing radiation pumping produces population
imbalances among the magnetic substates of the energy levels (that is,
atomic level polarization), in such a way that the populations of
substates with different values of $|M|$ are different. 
This is termed {\em atomic level alignment\/}.
As a result, the emission process can
generate linear polarization in spectral lines without the need for a
magnetic field. This is known as scattering line polarization \citep [e.g.,][]{Stenflo-1994,Landi-Landolfi-2004}. 
Moreover, radiation is also selectively absorbed
when the lower level of the transition is polarized \citep{Trujillo-Landi-1997,Trujillo-Bueno-1999,Trujillo-Bueno-2002, Manso-Trujillo-2003a}. Thus, the medium
becomes dichroic simply because the light itself has the chance of
escaping from it.

The fact that the illumination of the atoms in a stellar atmosphere is anisotropic is easy to understand if we consider the case of a plasma structure embedded in the optically thin outer layers (e.g., a solar coronal filament), because the incident radiation comes mainly from the underlying quiet photosphere and is contained within a cone of half aperture $\alpha{\le}90^{\circ}$, with the vertex centered on the point under consideration. The larger the height above the visible stellar ``surface'' the smaller $\alpha$ and the larger the anisotropy factor $w=\sqrt{2}J^2_0/J^0_0$, where $J^0_0{=}\oint \frac{{\rm d} \vec{\Omega}}{4\pi}I_{\nu, \vec{\Omega}}\,$ is the familiar mean intensity and 
${J}^2_0{\approx}\oint \frac{{\rm d} \vec{\Omega}}{4\pi}\frac{1}{2\sqrt{2}} (3\mu^2-1){{I_{\nu, \vec{\Omega}}}}\,$, with $\mu={\rm cos}{\theta}$ being the cosine of the angle between the ray and the stellar radius vector (hereafter, the vertical direction). Neglecting the $\vec{\Omega}$ dependence of the incident intensity $I_{\nu, \vec{\Omega}}$, and assuming that it is unpolarized, it is easy to find that $w=[1+{\rm cos}{\alpha}]{\rm cos}{\alpha}/2$, which shows that in this case $0{\le}w{\le}1$, with $w=1$ for the limiting case of a unidirectional unpolarized light beam that propagates along the vertical direction. 

The radiation field is also anisotropic within a stellar atmosphere itself (i.e., at heights where the overlying atmospheric plasma is not optically thin), but in this case $w$ can be positive or negative. As shown in the right panel of Fig. 5, at such heights the radiation propagating outwards shows limb darkening (i.e., it is predominantly {\em vertical}) while that propagating inwards shows limb brightening (i.e., it is predominantly {\em horizontal}). Therefore, there is competition, because ``vertical'' rays (i.e., with $|\mu|>1/{\sqrt{3}}$) make positive contributions to $w$, while ``horizontal'' rays (i.e., with $|\mu|<1/{\sqrt{3}}$) make negative contributions to $w$. As seen in Fig. 4 of \citep{Trujillo-Bueno-2001}, the larger the gradient of the source function the greater the anisotropy factor of the pumping radiation field, and the larger the amount of atomic level polarization. Therefore, at a given height in the inhomogeneous solar photosphere, the anisotropy factor of the solar {\em continuum radiation} in the (granular) upflowing regions is significantly larger than in the (intergranular) downflowing plasma \citep[see fig. 2 of][]{Trujillo-Bueno-2004}.

\begin{figure}
  \includegraphics[height=.3\textheight]{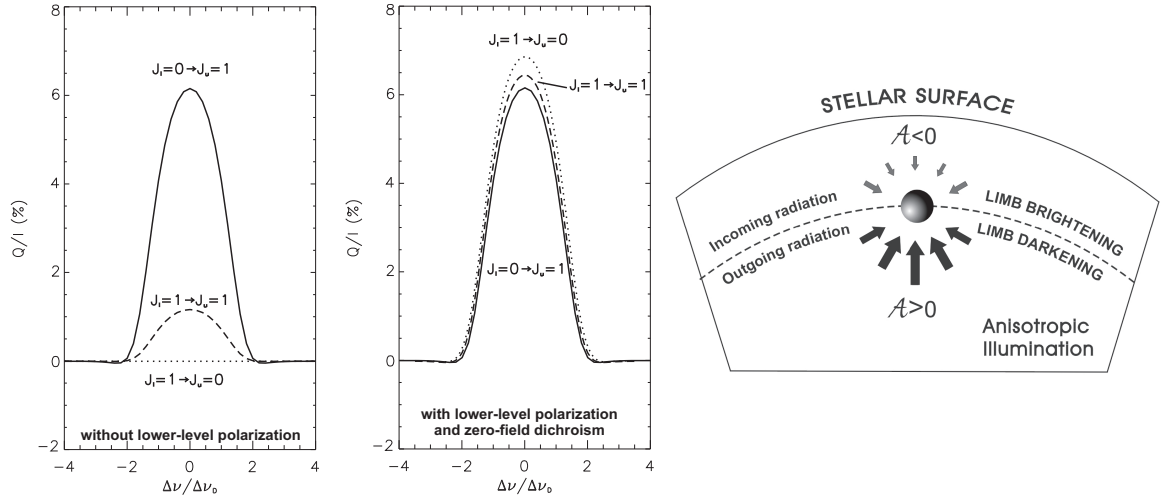}
  \caption{The emergent $Q/I$ profiles (for a line of sight with $\mu=0.1$) of three line transitions calculated in a model atmosphere with $T=6000$ K and $B=0$ G. All these $Q/I$ signals are solely due to the atomic level polarization that results from the anisotropic illumination illustrated in the right panel. Left panel: assuming that the lower level is unpolarized. Middle panel: taking into account the full impact of lower-level polarization. The reference direction for Stokes $Q$ is the parallel to the closest stellar limb. From \citep{Trujillo-Bueno-1999}.}
\end{figure}  

A useful expression to estimate the amplitude of the emergent fractional linear polarization is the following generalization of the Eddington-Barbier formula, which establishes that the emergent $Q/I$ at the line center of a sufficiently strong spectral line when observing along a line of sight specified by $\mu={\rm cos}{\theta}$ is approximately given by \citep[see][]{Trujillo-Bueno-2001}:

\begin{equation}
Q/I\,\approx\,{3\over{2\sqrt{2}}}(1-\mu^2)
[w_{J_uJ_\ell}^{(2)}\,\sigma^2_0({J_u})\,-\,w^{(2)}_{J_lJ_u}\,\sigma^2_0({J_l})],
\end{equation} 
where $w_{J_uJ_\ell}^{(2)}$ and $w^{(2)}_{J_lJ_u}$ are numerical factors that depend on the angular momentum values ($J$) of the lower ($l$) and upper ($u$) levels of the line transition (e.g., $w_{10}^{(2)} =1$, $w_{01}^{(2)} =0$ and $w_{11}^{(2)} =-1/2$), while $\sigma^2_0=\rho^2_0/\rho^0_0$ quantifies the {\em fractional atomic alignment} of the upper or lower level of the line transition under consideration\footnote{For example, $\rho^0_0(J=1)=(N_1+N_0+N_{-1})/\sqrt{3}$ and $\rho^2_0(J=1)=(N_1-2N_0+N_{-1})/\sqrt{6}$, where $N_1$, $N_0$ and $N_{-1}$ are the populations of the magnetic sublevels of a level with $J=1$.}. Note that the $\sigma^2_0$ values in Eq. (1) are those corresponding to the atmospheric height where the line center optical depth is unity along the line of sight.

Consider the three line transitions of Fig. 5, and the corresponding emergent $Q/I$ profiles obtained by solving numerically the scattering polarization problem in an unmagnetized model atmosphere assuming a two-level atomic model for each line independently. The left panel corresponds to calculations carried out assuming that the lower level is completely unpolarized, while the middle panel takes into account the full impact of lower-level polarization. Note that when the atomic polarization of the lower level is taken into account then the ``null line'' (i.e., the one which has unpolarized emission because it has $J_{u}=0$) shows the largest $Q/I$ amplitude. Isn't it fascinating? I mean the fact that ``zero-field'' dichroism (i.e., differential absorption of polarization components in the absence of any significant Zeeman splitting) is a very efficient mechanism for producing linear polarization in the spectral lines that originate in a stellar atmosphere.

\subsection{The Hanle effect}

In order to understand what the Hanle effect is it is first necessary to clarify that in the general case where polarization phenomena are taken into account, the full description of an atomic system requires us to specify, for each $J$-level, a density matrix with $(2J+1)^2$ elements\footnote{If the atom under consideration has non-zero nuclear angular momentum then a rigorous modeling of the spectral line polarization requires taking into account the hyperfine structure, which drastically increases the number of density matrix elements required to specify the excitation state.}. The diagonal ones, $\rho_{J}(M,M)$, quantify the populations of the individual sublevels and the non-diagonal ones, $\rho_{J}(M,M')$, the quantum coherences between each pair of them. We say that the quantum coherence $\rho_{J}(M,M')$ is non-zero when the atomic wave function presents a well defined phase relationship between the pure quantum states ${\vert}JM\rangle$ and ${\vert}JM^{'}\rangle$. The law of transformation of the density-matrix under a rotation of the reference system chosen for the specification of its elements indicates that it is actually very common to find non-zero coherences when describing the excitation state of an atomic system under the influence of anisotropic radiative pumping 
\citep[e.g., see][]{Landi-Landolfi-2004}.

Consider a reference system whose $Z$-axis (the quantization direction of total angular momentum) is chosen along the direction of the applied magnetic field and $J$-levels whose sublevels are not affected by possible crossings and/or repulsions with the sublevels pertaining to other levels. In this simplest case, the Hanle effect tends to reduce and dephase the quantum coherences with respect to the non-magnetic case, without modifying the population imbalances. For the Hanle effect to operate the magnetic field must be inclined with respect to the symmetry axis of the pumping radiation field. What happens is that as the sublevels are split by the magnetic field, the degeneracy of the $J$-level under consideration is lifted and the quantum coherences are modified. This gives rise to a characteristic magnetic-field dependence of the linear polarization of the emergent spectral line radiation that provides an attractive  diagnostic tool of magnetic fields in astrophysics \citep[e.g., the review by][]{Trujillo-Bueno-2001}.

Typically, in $90^{\circ}$ scattering geometry (e.g., when observing off the solar limb)
the largest polarization occurs for the unmagnetized case, with the direction of the linear polarization perpendicular to the scattering plane. In the presence of a magnetic field pointing either towards the observer (case ${\bf a}$) or away from him/her (case ${\bf b}$) the polarization amplitude is significantly reduced with respect to the unmagnetized case. Moreover, the direction of the linear polarization is rotated with respect to the zero field case. Normally, this rotation is counterclockwise for case (${\bf a}$) and clockwise for case (${\bf b}$). Therefore, when opposite magnetic polarities coexist within the spatio-temporal resolution element of the observation the direction of the linear polarization is like in the unmagnetized reference case, simply because the rotation effect cancels out. However, the polarization amplitude is indeed reduced with respect to the zero field reference case, which provides an ``observable'' that can be used to obtain information on hidden, tangled magnetic fields at subresolution scales in the solar atmosphere 
\citep[see][]{Stenflo-1994,Trujillo-Bueno-2004}.

On the other hand, in forward scattering geometry (e.g., when the line of sight points to solar disk center) we normally have zero polarization in the absence of magnetic fields, while the largest linear polarization (oriented parallel or perpendicular to the direction of the magnetic field) is found for a horizontal field with a strength such that the ensuing Zeeman splitting is sensibly larger than the level's natural width. In forward scattering geometry the linear polarization is created by the Hanle effect, a physical phenomenon that has been demonstrated to operate on the Sun via spectropolarimetric observations of coronal filaments in the He~{\sc i} 10830 \AA\ multiplet \citep[see][]{Trujillo-Bueno-2002}. Another very interesting example is shown in the right panel of Figure 4.

Approximately, the amplitude of the emergent
spectral line polarization is sensitive to magnetic strengths between
$0.1\,B_{\rm H}$ and $10\,B_{\rm H}$, where the critical Hanle field intensity ($B_{\rm
H}$, in gauss) is that for which the Zeeman splitting of the $J$-level
under consideration is equal to its natural width:

\begin{equation} 
B_{\rm H} = 1.137{\times}10^{-7}/(t_{\rm life}\,g_J)
\end{equation}
with $t_{\rm life}$ the lifetime, in seconds, of the $J$-level under
consideration and $g_J$ its Land\'e factor. If the line's lower level is the ground level or a metastable level, its ${t_{\rm life}(J_l)}\,{\approx}\,1/(B_{lu}J^0_0)\,{\approx}\,1/(\bar{n}A_{ul})]$ (with $\bar{n}=c^2J^0_0/(2h\nu^3)$ being the solid-angle average of the number of photons per mode at the frequency of the line under consideration and $B_{lu}$ and $A_{ul}$ the Einstein's coefficients). For relatively strong solar spectral lines ${t_{\rm life}(J_l)}$ 
is typically between a factor $10^2$ and $10^3$ larger than the upper-level lifetime (${t_{\rm life}(J_u)}{\approx}1/A_{ul}$, where $A_{ul}$ is Einstein's coefficient for the spontaneous emission process). For this reason, in solar-like atmospheres {\em the lower-level Hanle effect} is normally sensitive to magnetic fields in the milligauss range, while {\em the upper-level Hanle effect} is sensitive to fields in the gauss range.

\section{The non-LTE problem of the 2nd kind}

The generation and transfer of polarized radiation in a (generally magnetized) astrophysical plasma
is a very involved {\em non-local} and {\em non-linear} RT problem 
which requires jointly solving the statistical equilibrium equations for the elements
of the atomic density matrix and the Stokes-vector transfer equation
for each of the allowed transitions in the chosen multilevel model.
Once the selfconsistent values of the elements of the atomic density matrix are known at each point within the medium, it is straightforward to solve the Stokes vector transfer equation in order to obtain the emergent Stokes profiles. At present, there are two computer programs for solving non-LTE problems of the 2nd kind \citep[see][]{Manso-Trujillo-2003b,Stepan-2008}. Both codes are based on the formal solvers and the iterative methods proposed by Trujillo Bueno for the numerical solution of non-LTE polarization transfer problems \citep[see][]{Trujillo-Bueno-1999,Trujillo-Bueno-2003}.

\subsection{The Stokes-vector transfer equation}

In the polarized case, instead of the standard
RT equation for the specific intensity ${I}({\nu},{\vec{\Omega}})$
one has to solve, in general, 
the following {\em vectorial} transfer equation
for the Stokes vector ${\bf I}({\nu},{{\vec{\Omega}}})=({I,Q,U,V})^{\dag}$
(${\dag}\,=\,{\rm transpose}$):

\begin{equation}
{{d}\over{ds}}{\bf I}\,=\,{\boldmath {\epsilon}} - {{\bf K}}{\bf I},
\end{equation}
where $s$ measures the geometrical distance 
along the ray of direction ${\vec{\Omega}}$,
{\boldmath ${\epsilon}$} is the emission vector, and $\bf K$ the propagation matrix.
The elements of the emission vector (that is, 
$\epsilon_I$, $\epsilon_Q$, $\epsilon_U$, and $\epsilon_V$) account for the contribution of the
{\em spontaneous emission} process to the 
intensity and polarization that is generated at each spatial point 
by the physical mechanisms under consideration. The propagation matrix $\bf K$ can be written as

\begin{eqnarray}
\left( \begin{array}{cccc}
{\eta_I}&{0}&{0}&{0} \\
{0}&{\eta_I}&{0}&{0} \\
{0}&{0}&{\eta_I}&{0} \\
{0}&{0}&{0}&{\eta_I} 
\end{array} \right)\,+\,
\left( \begin{array}{cccc}
{0}&{\eta_Q}&{\eta_U}&{\eta_V} \\
{\eta_Q}&{0}&{0}&{0} \\
{\eta_U}&{0}&{0}&{0} \\
{\eta_V}&{0}&{0}&{0} 
\end{array} \right)\,+\,
\left( \begin{array}{cccc}
{0}&{0}&{0}&{0} \\
{0}&{0}&{\rho_V}&{-\rho_U} \\
{0}&{-\rho_V}&{0}&{\rho_Q} \\
{0}&{\rho_U}&{-\rho_Q}&{0} 
\end{array} \right),
\end{eqnarray}
which helps to clarify that it has three contributions:
absorption (the first matrix, $\bf K_1$, which is responsible for the
attenuation of the radiation beam irrespective of its polarization state), 
dichroism (the second matrix, $\bf K_2$, which accounts for a selective
absorption of the different polarization states),
and dispersion (the third matrix, $\bf K_3$, which describes the dephasing of the different
polarization states as the radiation beam propagates through the medium).
Actually, the situation is more complicated (and interesting!) because
in general the {\em propagation} matrix is the difference of two matrices: 
${\bf K}={\bf K}^{\rm A}-{\bf K}^{\rm S}$, where ${\bf K}^{\rm A}$ is the above-mentioned contribution
resulting from transitions from the lower-level towards the upper-level, while
${\bf K}^{\rm S}$ is the contribution caused by stimulated emission processes.
Thus, ${\bf K}_1^{\rm S}$ would be the amplification matrix (responsible 
for the amplification of the radiation beam irrespective of its polarization state),
${\bf K}_2^{\rm S}$ would be the dichroism matrix (responsible of a selective
stimulated emission of different polarization states), and ${\bf K}_3^{\rm S}$
would be the corresponding dispersion matrix. Taking ${\bf K}_2^{\rm S}$
and ${\bf K}_3^{\rm S}$ into account may be of 
great interest for new discoveries in the field of 
astronomical masers \citep[e.g.,][]{Asensio-etal-2005}.

The general expressions of the components of the emission vector
and of the propagation matrix are
very involved and will not be written here \citep[see Chapter 7 of][]{Landi-Landolfi-2004}.
They are given in terms of the 
$\rho^K_Q$ multipolar components of the atomic density matrix ($K=0,1,2$ and $-K{\le}Q{\le}K$) 
corresponding to the upper and lower levels of the
line transition under consideration and of line-shape profiles, whose dependence
on the magnetic quantum numbers cannot be neglected when the Zeeman splittings
are a significant fraction of the spectral line width. Fortunately, such general expressions simplify
considerably for several cases of practical interest . For instance, the following are the 
expressions of the only non-zero emission coefficients for the case of a one-dimensional atmosphere, either unmagnetized, or permeated by a {\em microturbulent} and {\em isotropically distributed}
magnetic field of strength $B$, or in the presence of a magnetic field with a fixed orientation but with a strength in the saturation regime of the Hanle effect\footnote{In this regime the quantum coherences turn out to vanish
in a reference system whose quantization axis is parallel to the magnetic field vector
(i.e. the real and imaginary parts of the
$\rho^2_1$ and $\rho^2_2$ components are zero).}:

\begin{equation}
  \epsilon^{\rm line}_I\,=\,\epsilon_0 [\rho^0_0+
	w_{J_uJ_\ell}^{(2)} \frac{1}{2\sqrt{2}} (3\mu^2-1)\rho^2_0],
\end{equation}
\begin{equation}
  \epsilon^{\rm line}_Q\,=\,\epsilon_0 w_{J_uJ_\ell}^{(2)} \frac{3}{2\sqrt{2}}(1-\mu^2)
	\rho^2_0,
\end{equation}
where the $\rho^K_0$ ($K=0,2$)
values are those of the {\it upper} level of the line transition under consideration,
$\epsilon_0=(h\nu/4\pi)A_{u\ell}{\phi_x}{({2J_u+1})^{1/2}}{\cal N}$ 
(with $h$, ${\cal N}$ and $\phi_x$ being, respectively, the Planck constant, the particle density and 
the normalized line profile, with $x$ the frequency distance from the 
line center in units of the Doppler width), 
$w_{J_uJ_\ell}^{(2)}$ is the previously mentioned numerical factor that only depends on the 
quantum numbers of the levels involved in the transition,
and where the orientation of the ray is specified by the azimuthal angle $\chi$ and by 
$\mu={\rm cos}\theta$ (with $\theta$ the angle between the direction of the radiation beam 
and the quantization axis, which must be the magnetic field direction itself for the deterministic field case 
and the normal to the surface of the stellar 
atmospheric model for the two other cases). Likewise, the only non-zero elements of the propagation matrix are
$\eta_I$ and $\eta_Q$ which are given by
identical expressions (i.e. by $\eta_I=\epsilon_I$ and 
$\eta_Q=\epsilon_Q$),
but with $\eta_0=(h\nu/4\pi)B_{lu}{\phi_x}{({2J_l+1})^{1/2}}{\cal N}$ instead of $\epsilon_0$,
$w^{(2)}_{J_lJ_u}$ instead of $w^{(2)}_{J_uJ_l}$ and with the $\rho^K_0$
values of the {\it lower} level of the line transition instead of those of the upper level
(for the case in which stimulated emissions are neglected).
Note that $\epsilon_I$ and $\eta_I$
depend on both the level's overall population ($\rho^0_0$) and the population imbalances ($\rho^2_0$), 
while $\epsilon_Q$ and $\eta_Q$
depend only on the population imbalances ($\rho^2_0$)\footnote{In these expressions,
the reference direction for $Q>0$ is the perpendicular to the plane formed by the
propagation direction and the Z-axis.}.

Under such assumptions
a very useful formula can be derived to estimate the emergent
fractional linear polarization amplitude at the center of a spectral line:

\begin{equation}
   {\frac{Q}{I}}\,{\approx}\,
	\frac{\epsilon_Q/\epsilon_I-\eta_Q/\eta_I}
	{1-(\eta_Q/\eta_I)(\epsilon_Q/\epsilon_I)},
\end{equation}
where $\epsilon_I$, $\epsilon_Q$, $\eta_I$ and $\eta_Q$ 
must be evaluated at height in the model atmosphere where the
line-center optical depth is unity along the line of sight.
This expression can be simplified taking into account 
the low polarization level in solar-like atmospheres (i.e., that $\eta_Q/\eta_I\ll 1$ and 
$\epsilon_Q/\epsilon_I\ll 1$). Thus, to first order in
$\epsilon_Q/\epsilon_I$ and $\eta_Q/\eta_I$, the fractional polarization
reduces to $Q/I\,{\approx}\,{\epsilon_Q}/{\epsilon_I}-{\eta_Q}/{\eta_I}$, 
which leads directly to the approximate Eq. (1). Its first term is due to {\em selective emission} of polarization components (caused by the population imbalances of the upper level), while its second term is due to ``zero-field'' dichroism (i.e., it is caused by the {\em selective absorption} of polarization components produced by the population imbalances of the lower level). Eq. (1) is very useful for understanding that, in general, the second solar spectrum is produced by such two contributions \citep[see][]{Trujillo-Bueno-2009a}. 

\subsection{The rate equations for the elements of the atomic density matrix}

We need also to consider the master equation for the atomic density matrix
\citep[see][]{Cohen-1992,Landi-Landolfi-2004}.
Perhaps, the simplest and most famous example of this approach is that formulated
by Einstein, who in 1917 introduced rate equations describing the effect
of absorption, stimulated emission, and spontaneous emission processes
in a two-level atom. The situation is considerably 
more complex (and interesting!) in the present polarized case
in which we have {\em atomic and light polarization} instead of simply
overall populations of the atomic levels and intensity. In general, we have:

\begin{equation}
{{\rm d}\over{\rm dt}}{\rho^K_Q(J)}\,=\,{[}{{\rm d}\over{\rm dt}}{\rho^K_Q(J)}{]}_{\rm Magnetic\,Fields}\,+\,\,{[}{{\rm d}\over{\rm dt}}{\rho^K_Q(J)}{]}_{\rm Radiation}\,+\,\,{[}{{\rm d}\over{\rm dt}}{\rho^K_Q(J)}{]}_{\rm Collisions},
\end{equation}
where the first two terms on the RHS are the contributions 
due to a magnetic field (e.g., via the Hanle effect) 
and to radiative transitions, while the last term is that due to collisions. 
In order to calculate the $\rho^K_Q$ unknowns at each grid point of the chosen atmospheric model 
one usually assumes statistical equilibrium (i.e., ${{\rm d}{\rho^K_Q(J)}/{\rm dt}}\,=\,0$).

The quantum theory of spectral line polarization considered here assumes that the absorption and re-emission processes are statistically independent events. When the only significant quantum coherences are those between the sublevels pertaining to each degenerate level, this  
complete redistribution theory \citep[see][]{Landi-Landolfi-2004} provides a suitable description 
if the pumping radiation has no spectral structure across a frequency interval larger than both the Larmor frequency (see below) and the natural width of the atomic levels. 
 
Of great interest is the Hanle-effect contribution of Eq. (8), which in the magnetic field reference frame 
reads \citep[e.g., Eq. 7.11 in][]{Landi-Landolfi-2004}

\begin{equation}
{[}{{\rm d}\over{\rm dt}}{\rho^K_Q(J)}{]}_{\rm Hanle}\,{=}\,-2{\pi}i\,\nu_{\rm L}\,{g_J}\,Q\,{\rho^K_{Q}(J)},
\end{equation}
where the Larmor frequency ${\nu}_{\rm L}=1.3996{\times}10^6B$ (with $B$ expressed in gauss) and $g_{J}$ is the Land\'e factor of the $J$-level under consideration. This expression, which is valid for the case of a multilevel atom assuming that quantum coherences can exist {\em only} between the magnetic sublevels pertaining to each individual $J$-level, 
indicates that {\em in the magnetic field reference system} the population imbalances (i.e. the $\rho^K_Q$ density-matrix elements with $Q=0$)
are insensitive to the intensity of the magnetic field, while the coherences 
(i.e. the $\rho^K_Q$ elements with $Q{\ne}0$) are reduced and dephased as the magnetic field strength increases.
In fact, when the Zeeman splitting is much larger than the inverse lifetime of the shortest lived levels
(e.g., when $B>B_{\rm satur}(J_u)\,{\approx}\,10\,B_H(J_u)$) such coherences vanish in the magnetic field reference system.

Certainly isotropic collisions with electrons and ions contribute to the rate of change of the ${\rho^K_Q(J)}$ elements, but the most relevant effect in solar-like atmospheres comes typically from isotropic {\em elastic}  collisions with neutral hydrogen atoms which induce transitions between the magnetic sublevels pertaining to the $J$-level under consideration. It can be shown that for isotropic elastic collisions one has \citep[e.g., see  Eq. 7.101 in][]{Landi-Landolfi-2004}

\begin{equation}
{[}{{\rm d}\over{\rm dt}}{\rho^K_Q(J)}{]}_{\rm Elastic\,Collisions}\,{=}\,-D^{(K)}(J)\,{\rho^K_{Q}(J)},
\end{equation}   
where $D^{(K)}(J)$ is the depolarizing rate (in ${\rm s}^{-1}$) whose value is proportional to the neutral hydrogen number density. Obviously $D^{(0)}(J)=0$, and the effect on the $\rho^K_{Q}(J)$ elements with $K>0$ 
will be significant when $\delta^{(K)}(J)=D^{K}(J)\,t_{\rm life}(J)\,{\approx}\,1$, 
where $t_{\rm life}(J)$ is that defined after Eq. (2).

Finally, the contribution caused by radiative transitions includes the {\em transfer}
rates due to absorption ($T_A$), spontaneous emission ($T_E$) and stimulated
emission ($T_S$) from other levels, and the {\em relaxation} rates due to absorption ($R_A$), 
spontaneous emission ($R_E$) and stimulated emission ($R_S$)
towards other levels. The general expressions are rather 
complicated \citep[see, e.g., \S7.2.a of][]{Landi-Landolfi-2004}, but 
the following ones are of great practical interest even though they
neglect the stimulated emission rates and correspond to the particular case of a one-dimensional atmosphere, either unmagnetized or in the presence of a deterministic magnetic field with a fixed orientation but with a strength in the saturation regime of the Hanle effect\footnote{It is important to note that in this last case 
the multipolar components, ${\rho^K_Q(J)}$, and the radiation field tensors,
${{\bar J}^K_Q}$, have to be defined in the magnetic field reference frame.}:

\begin{eqnarray} 
\big[\frac{d}{dt}\; \rho_0^{K} (J)\big]_{\rm Radiation} & = &  
\sum_{J_l}\sum_{K_{l}}\,{{\rho^{K_{l}}_{0}(J_l)}}\,
{T_{\rm A}}(J_l;K_{l} \rightarrow J;K) \nonumber \\ &&
+\,\sum_{J_u}\sum_{K_{u}}\,{{\rho^{K_{u}}_{0}(J_u)}}\,
{T_{\rm E}}(J_u;K_{u}\rightarrow J;K) \nonumber \\ &&
-\,{\sum_{K^{'}}}\,{{\rho^{K^{'}}_{0}(J)}}\,\big[
{R_{\rm A}}(J;K,K^{'}\rightarrow J_u)
+\,{R_{\rm E}}(J;K,K^{'} \rightarrow J_l) \big], 
\end{eqnarray}
where $K_l=0,2,4,...,2J_l$; $K_u=0,2,4,...,2J_u$ and $K^{'}=0,2,4,...,2J$.

Of particular interest are the transfer rate $T_{\rm A}$ due to {\em absorption}
from lower levels and the relaxation rate $R_{\rm A}$ due to absorptions toward
upper levels. Their expressions are given by linear combinations of the ${{\bar J}^0_0}$ and 
${{\bar J}^2_0}$ radiation field tensors \citep[see Eqs. 7.14a and 7.14d in][]{Landi-Landolfi-2004}. 
Note that ${{\bar J}^0_0}$ is the familar ${\bar J}$-quantity of the standard
non-LTE problem of the $1st$ kind (which is an average over frequencies and directions of the Stokes
$I$ parameter weighted by the absorption profile), while\footnote{In the most general case the number of ${{\bar J}^K_Q}$ tensors necessary to quantify the symmetry properties of the radiation field is 9 \citep[see \S5.11 in][]{Landi-Landolfi-2004} \citep[see also Fig. 3 in][]{Trujillo-Shchukina-2009}.}
 
\begin{equation}
{{{\bar J}^2_0}}=\int {\rm d}x 
\oint \phi_x \frac{{\rm d} \vec{\Omega}}{4\pi}
\frac{1}{2\sqrt{2}} \left[(3\mu^2-1){{I_{x \vec{\Omega}}}}+
3(\mu^2-1){Q_{x \vec{\Omega}}}\right]\,.
\end{equation}

\section{An interesting effect: magnetic modulation of the population ratio}
 
Equations (11) imply that even the ratio ${\cal N}_{u}/{\cal N}_{l}$ of the  
populations of the upper and lower levels 
of a spectral line may depend on the ${{\bar J}^2_0}$ tensor component (which quantifies the ``anisotropy'' of the radiation field), and not only on ${{\bar J}^0_0}$. Even more interesting is the fact that the ${\cal N}_{u}/{\cal N}_{l}$ value is modulated by the inclination of the magnetic field with respect to the symmetry axis of the incident radiation field (hereafter, the vertical axis). In order to illustrate this remarkable phenomenon let us consider the case of anisotropic radiation pumping 
in a gas composed by two-level atoms with $J_l=1$ and $J_u=0$ in the presence of a deterministic magnetic field inclined by an angle $\theta_B$ with respect to the vertical axis and with a strength $B>B_{\rm satur}(J_l){\approx}10\,B_H(J_l)$, so that all the quantum coherences between the magnetic sublevels of the lower level are zero in the magnetic field reference frame. In this Hanle effect saturation regime the statistical equilibrium equations in the magnetic field reference frame are identical to those corresponding to the unmagnetized case \citep[see Eqs. 3--5 in][]{Trujillo-Landi-1997}, with the only difference that ${{\bar J}^2_0}$ has to be substituted by:

\begin{equation}
[{{\bar J}^2_0}]_{\rm {B}}\,=\,{1\over{2}}(3{\rm cos}^2{\theta_B}-1)\,[{{\bar J}^2_0}]_{\rm {V}},
\end{equation}
where $[{{\bar J}^2_0}]_{\rm {B}}$ and $[{{\bar J}^2_0}]_{\rm {V}}$ are the ${{\bar J}^2_0}$ radiation field tensor in the magnetic field and in the ``vertical'' reference frames, respectively. Neglecting the contribution of inelastic collisions in Eq. (3) of \citep{Trujillo-Landi-1997} one obtains

\begin{equation}
{{\cal N}_{u}\over{{\cal N}_{l}}}={\bar{n}\over{3}}\,\big{(}\,1\,-\,{1\over{\sqrt{2}}}\,[w]_{B}\,
[\sigma^2_0(J_l)]_{B}\,\big{)},
\end{equation}
where $\bar{n}={{\bar J}^0_0}(2J_l+1)B_{lu}/(2J_u+1)A_{ul}$ 
is the average of the number of photons per mode of the incident 
radiation at the line frequency  
and $[w]_{B}=\sqrt{2}[{{\bar J}^2_0}]_{\rm {B}}/{{\bar J}^0_0}$ is the anisotropy factor in the magnetic field reference frame. In Eq. (14) $[\sigma^2_0(J_l)]_B=[\rho^2_0(l)]_B/\rho^0_0(l)$ is the fractional atomic alignment of the lower-level in the magnetic field reference frame, which can be easily calculated by neglecting inelastic collisions in Eq. (4) of \citep{Trujillo-Landi-1997}. Disregarding also for simplicity elastic collisions and using the ensuing expression in Eq. (14) it is easy to find that for a two-level atom with $J_l=1$ and $J_u=0$

\begin{equation}
{{\cal N}_{u}\over{{\cal N}_{l}}}={\bar{n}\over{3}} \, \big{[}\,\,1\,-\,{ { { (3{\rm cos}^2{\theta_B}-1) }^{2} w^2/2 }
\over{4\,-\,(3{\rm cos}^2{\theta_B}-1)\,w}}\,\,\big{]}.
\end{equation}
The lower curve in Fig. 6 shows the variation with $\theta_B$ of ${\cal N}_{u}/{\cal N}_{l}$ for this case
of a resonance line transition with $J_l=1$ and $J_u=0$, assuming $\bar{n}=10^{-4}$ (which implies
that the rates of stimulated emission can indeed be neglected) and $w=0.99$ (i.e., that the atomic system is irradiated by an almost collimated radiation beam)\footnote{It is of interest to point out that such values for $\bar{n}$ and $w$ are very similar to those corresponding to the 10830 \AA\ solar continuum radiation at a distance of 10 solar radii above the visible stellar surface.}. Note that when $w=1$ the upper level  
is fully depopulated if the magnetic field is parallel to the incident beam of unpolarized radiation (i.e., if 
$\theta_B=0^{\circ}$ or $\theta_B=180^{\circ}$). This occurs because, under such anisotropic pumping conditions, 
all atoms end up in the $M_l=0$ sublevel of the lower level and the medium becomes transparent.
The same happens in the absence of magnetic fields, a well-known effect in the quantum optics literature \citep[e.g.,][]{Happer-1972}.   

A similar but more involved analytical expression for ${\cal N}_{u}/{\cal N}_{l}$
can also be derived for the case of a resonance line transition with $J_l=J_u=1$, but assuming now that
$B>B_{\rm satur}(J_u){\approx}10\,B_H(J_u)$.
To this end it suffices with using the statistical equilibrium equations 
for the case of a two-level atom with $J_l=J_u=1$ \citep[see][]{Trujillo-Bueno-1999,Trujillo-Bueno-2003}.  
The upper curve in Fig. 6 shows the ensuing results. Note that for this case 
${\cal N}_{u}/{\cal N}_{l}=\bar{n}=10^{-4}$ when $[w]_{B}=0$ (which is always the case at $\theta_B=54.74^{\circ}$ and $\theta_B=125.26^{\circ}$, the so-called Van-Vleck angles).  

As seen in Fig. 6, for both line transitions the population ratio ${\cal N}_{u}/{\cal N}_{l}$ depends on the inclination of the magnetic field with respect to the symmetry axis of the pumping radiation field. Note that for the operation of this interesting effect in astrophysical environments where the stimulated emission terms are negligible (i.e., because $\bar{n}{\ll}1$) the lower level of the line transition under consideration must be polarized. However, if stimulated emission radiation is significant (i.e., because $\bar{n}\,{\sim}\,1$) then such a modulation takes place even in line transitions whose lower-level is unpolarized, such as in a spectral line with $J_l=0$ and $J_u=1$. 

\begin{figure}
  \includegraphics[height=.3\textheight]{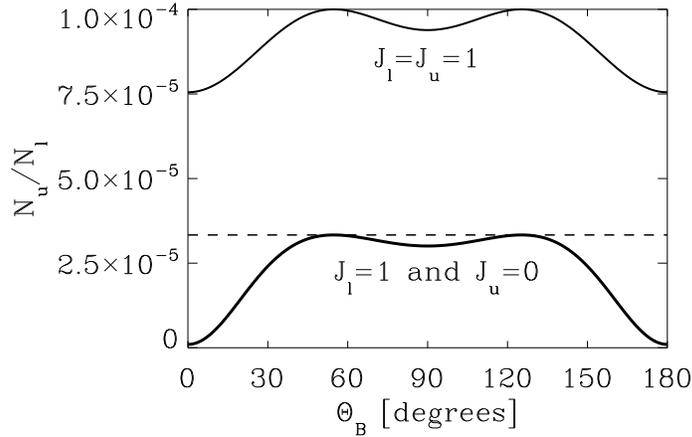}
  \caption{Variation of the indicated population ratio with the inclination of the magnetic field vector with respect to the symmetry axis of the incident radiation. This radiation has $\bar{n}=10^{-4}$ and $w=[w]_{V}=0.99$. The strength of the magnetic field is assumed to be sufficiently high so as to produce saturation of the Hanle effect for the spectral line under consideration (e.g., only 0.1 G for the line with $J_u=0$ and 100 G for the line with $J_l=J_u=1$). The dashed line indicates the $\bar{n}/3$ value. }
\end{figure} 

\section{Concluding comments}

As mentioned in \S1, we may define {\em the Sun's hidden magnetism} as all 
the magnetic fields of the extended solar atmosphere that are impossible 
to diagnose via the consideration of the Zeeman effect alone. 
Contrary to what one might think, 
there are many examples that belong to this category:

\begin{itemize}

\item The greater part of the magnetism of the quiet solar photosphere.

\item The magnetic fields of the solar chromosphere outside sunspots. 

\item The magnetic fields that confine the plasma of solar prominences and 
filaments.

\item The magnetism of the solar transition region and corona.

\end{itemize}

Fortunately, the Hanle and Zeeman effects can be suitably complemented for exploring magnetic fields in solar and stellar physics. To this end, it is necessary to interpret high-sensitivity spectropolarimetric observations through the numerical solution of non-LTE problems of the 2nd kind. The interested reader will find abundant information on a variety of interesting applications in solar physics in the following review papers: \citep{Trujillo-Bueno-2006,Casini-2007,Trujillo-Bueno-2009a,Trujillo-Bueno-2009b}.

Finally, it is important to emphasize that one 
of the greatest challenges in astrophysics is the empirical investigation of the magnetic field vector in a variety of astrophysical systems, such as the solar corona, circumstellar envelopes, accreting systems, etc.
The physical mechanisms I have considered here (anisotropic radiative pumping, atomic level polarization and the Hanle and Zeeman effects) occur in many astrophysical environments, not only in the solar atmosphere. In particular, the linear polarization signals produced by the presence of atomic level polarization 
are sensitive to magnetic fields in a parameter domain that ranges from very weak fields (e.g., 1 mG) to a few hundred gauss. The observation and physical interpretation of these polarization effects via the solution of the non-LTE problem of the 2nd kind provides key information, impossible to obtain via spectroscopy or conventional spectropolarimetry.

%%%%%%%%%%%%%%%%%%%%%%%%%%%%%%%%%%%%%%%%%%%%%%%%
%% BACKMATTER
%%%%%%%%%%%%%%%%%%%%%%%%%%%%%%%%%%%%%%%%%%%%%%%%

\begin{theacknowledgments}
  Financial support by the Spanish Ministry of
  Science and Innovation through project AYA2007-63881 and by the European Commission
  via the SOLAIRE network (MTRN-CT-2006-035484) are gratefully acknowledged. 
  The author is also grateful to the members of the SOC for their invitation to participate in a very good 
  conference. 
\end{theacknowledgments}



\bibliographystyle{aipproc}   % if natbib is available

%\bibliography{sample}

%  \typeout{******************************************}
%  \typeout{** Please run "bibtex \jobname" to optain}
%  \typeout{** the bibliography and then re-run LaTeX}
%  \typeout{** twice to fix the references!}
%  \typeout{******************************************}

%%%%\end{document}

%%%%%%%%%%%%%%%%%%%%%%%%%%%%%%%%%%%%%%%%%%%
%% The following lines show an example how to produce a bibliography
%% without the help of the BibTeX program. This could be used instead
%% of the above.
%%%%%%%%%%%%%%%%%%%%%%%%%%%%%%%%%%%%%%%%%%%

\end{document}